\documentclass{appolb}
\usepackage{graphicx}
\usepackage{url}
\usepackage{cite}
\begin{document}
\title{Development of the (d,n) proton-transfer reaction\\ in inverse kinematics for structure studies%
\thanks{Presented at the XXXV Mazurian Lakes Conference on Physics, Piaski, Poland, September
3-9, 2017}}%
\author{K.L.  Jones$^a$,
C.~Thornsberry$^a$, J.~Allen$^b$, A.~Atencio$^c$, D.W.~Bardayan$^b$,
D.~Blankstein$^b$, S.~Burcher$^a$, A.B.~Carter$^a$, K.A.~Chipps$^d$,
J.A.~Cizewski$^c$, I.~Cox$^a$, Z.~Elledge$^a$, M.~Febbraro$^d$, A. 
Fija\l{}kowska$^c$, R.~Grzywacz$^a$, M.R.~Hall$^b$, T.T.~King$^a$,
A.~Lepailleur$^c$, M.~Madurga$^a$, S.T.~Marley$^e$, P.D.~O'Malley$^b$,
S.V.~Paulauskas$^a$, S.D.~Pain$^d$, W.A.~Peters$^f$, C.~Reingold$^b$,
K.~Smith$^a$  \footnote{Current address: Los Alamos National Laboratory, Los Alamos, NM 87545, USA}, S.~Taylor$^a$, W.~Tan$^b$, M.~Vostinar$^a$,
D.~Walter$^c$
\address{$^a$Department of Physics and Astronomy, University of Tennessee, Knoxville, TN~37996, USA \\
$^b$Department of Physics, University of Notre Dame, Notre Dame, IN 46556, USA \\
$^c$Department of Physics and Astronomy, Rutgers University, New Brunswick, NJ~08903, USA\\
$^d$Physics Division, Oak Ridge National Laboratory, Oak Ridge, TN 37831, USA\\
$^e$Department. of Physics Louisiana State University, USA\\
$^f$Joint Institute for Nuclear Physics and Applications, Oak Ridge, TN 37831, USA 
}}

\maketitle
\begin{abstract}
Transfer reactions have provided exciting opportunities to study the
structure of exotic nuclei and are often used to inform studies relating to
nucleosynthesis and applications.  In order to benefit from these reactions
and their application to rare ion beams (RIBs) it is necessary to develop
the tools and techniques to perform and analyze the data from reactions
performed in inverse kinematics, that is with targets of light nuclei and
heavier beams.  We are continuing to expand the transfer reaction toolbox in
preparation for the next generation of facilities, such as the Facility for
Rare Ion Beams (FRIB), which is scheduled for completion in 2022.  An
important step in this process is to perform the (d,n) reaction in inverse
kinematics, with analyses that include Q-value spectra and differential
cross sections.  In this way, proton-transfer reactions can be placed on the
same level as the more commonly used neutron-transfer reactions, such as
(d,p), ($^9$Be,$^8$Be), and ($^{13}$C,$^{12}$C).  Here we present an
overview of the techniques used in (d,p) and (d,n), and some recent data
from (d,n) reactions in inverse kinematics using stable beams of $^{12}$C
and $^{16}$O.
\end{abstract}
  
\section{Introduction: Deuteron-induced transfer reactions}
\begin{minipage}{1.0\textwidth}
\vspace{5mm}
\begin{center}
{\small
{\em Deuteron} \\
Proton and neutron;\\
Tensor force brings together\\
fragile building block. \\}
\end{center}
\end{minipage}
\\
\vspace{5mm}

\begin{figure}[ht]
\centerline{%
\hspace{-2cm} \includegraphics[trim = 0 0 0 100, clip,width=11cm]{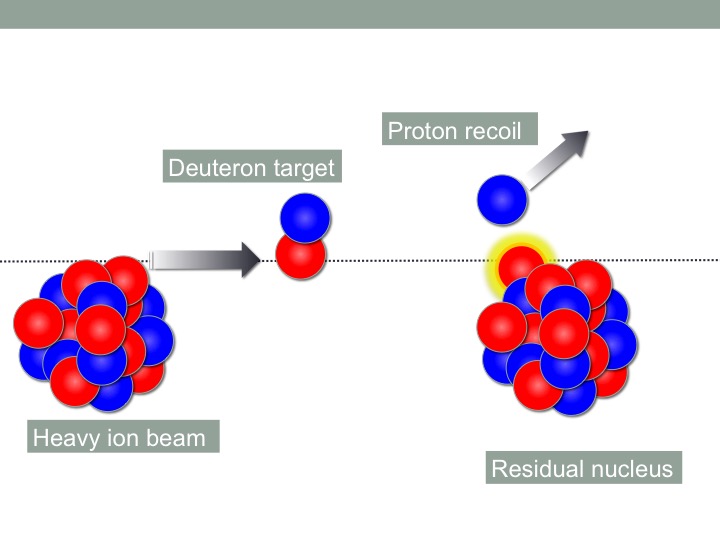}}
\caption{Cartoon of a (d,p) reaction in inverse kinematics.  A RIB impinges on a deuteron target resulting in a residual nucleus and a proton.  Adapted from
Ref.~\cite{Jon13} }
\label{Fig:cartoon}
\end{figure}

Transfer reactions can be used to elucidate the structure of nuclei through
measurements of the Q-value of the reaction, and the intensity and shape of
angular distributions of particles emerging from the reaction.  When rare
ion beams (RIBs) undergo transfer reactions on targets of light ions,
commonly referred to as inverse kinematics measurements (see Fig.
\ref{Fig:cartoon}), the structure of exotic nuclei can be studied in detail. 
The first neutron transfer reactions performed in inverse kinematics used
stable $^{132, 136}$Xe beams at the Gesellschaft f{\"u}r
Schwerionenforschung (GSI) \cite{Kra91}.  This pioneering experiment laid
the groundwork for many neutron transfer studies performed with RIBs since
the mid-1990s, in particular using the (d,p) single-neutron transfer
reaction.

The Q-value of a (d,p) reaction is extracted from the measured angle and
energy of the proton.  The conversion from the laboratory to the center-of-mass frame causes kinematic compression of the spectrum of states in the
residual nucleus.  However, in most cases the limiting factor in the
resolution is the CD$_2$ target thickness, and commonly targets of 100 -
200~$\mu$g/cm$^2$ are used in order to resolve individual states, while
still achieving statistical significance.  Thicker targets can be used when
states are identified from their $\gamma$ decays (see for example
Refs. \cite{Lab10,Sim00, Bil07,Ebe01,Dig11,Sve05}), but in this case the
statistical uncertainty is often limited by the efficiency of the $\gamma$-ray detector system.\\

The shapes of the angular distributions of protons emitted from (d,p)
reactions can be indicative of the $\ell$ transfer of the reaction.  The
sensitivity of the angular distributions to the transferred $\ell$ is
dependent on the beam energy, with very low or very high beam energies
(compared to the energy of the Coulomb barrier) producing fairly flat
angular distributions, regardless of $\ell$ transfer.

Another quantity of interest that can be extracted from transfer reactions
is the spectroscopic factor, $S$, which is connected to the structure of the
nucleus through the single-particle radial overlap function $u_{\ell s j}$
and the normalized wave function, $v_{\ell s j}$:
\begin{equation}
S_{\ell s j}=|{A_{\ell s j}}|^2
\end{equation}
where A$_{\ell s j}$ is the spectroscopic amplitude, and:
\begin{equation}
 u_{\ell s j}(r)= A_{\ell s j} v_{\ell s j}(r).
\end{equation}
However,  as $S$ is extracted from a measured normalized differential cross-section and an angular distribution calculated from a reaction model:
\begin{equation}
S_{exp}=\frac{d\sigma_{exp}/d\Omega}{d\sigma_{calc}/d\Omega}
\end{equation}
$S$ is not an observable of the experiment and is model dependent.  The main
sources of uncertainty in the calculations come from the optical
(scattering) potentials and the bound state potential in the final state. 
Optical potentials can be extracted by fitting elastic scattering data. 
Different optical potentials can produce angular distributions with both
different shapes and intensities.  As for the bound state of the residual
nucleus, this is typically modeled by a Woods-Saxon potential with the depth
adjusted to the binding energy of the state, and the geometry defined by the
radius and diffuseness.  At beam energies close to the Coulomb barrier the
magnitudes of the calculated differential cross sections can be very
sensitive to the bound-state potential used.

\section{The (d,n) single proton transfer reaction}

In stark contrast to the recent growth in inverse kinematics (d,p)
measurements, there have been fewer developments made for the (d,n)
single-proton transfer reaction, in particular where the nuclear structure
is extracted from measuring the emitted neutrons (see for example
\cite{Pet13,Feb17,Bab18}).  Our collaboration has been developing techniques
for performing the (d,n) proton transfer reaction with RIBs using the
Versatile Array of Neutron Detectors at Low Energy (VANDLE) \cite{Pau14,
Pet16}.  The energy and angle of protons emitted from (d,p) reactions can be
measured with relative ease in silicon detectors.  It is less
straightforward to measure the energy of neutrons emerging from (d,n)
reactions.  VANDLE uses the time of flight of neutrons to find their kinetic
energy.  However, the plastic scintillator from which VANDLE is built is
also sensitive to $\gamma$ rays, which can cause a limiting source of
background in RIB experiments.  \\

Our first attempt to measure states in $^8$B using the $^7$Be(d,n)$^8$B
reaction, at the TWINSOL facility at the University of Notre Dame
\cite{Bec16}, suffered from excessive levels of background in the neutron
spectra.  This was due in part to the configuration where the VANDLE bars
were in the direct path of neutrons and $\gamma$~rays emitted from the RIB
source in TWINSOL.  A measurement made in the same experimental campaign
utilizing the $^{17}$F(d,n)$^{18}$Ne reaction produced results from
deuterated liquid scintillator detectors \cite{Feb14} where pulse-shape
discrimination (PSD) was used to separate neutrons from $\gamma$~rays.  This
largely eliminated the background issues.  However, a limitation in this
technique is the requirement that neutrons deposit energy above a threshold
of about 100~keVee before PSD separation can be achieved.  For comparison,
in the analysis presented below the threshold for VANDLE was set at
30~keVee, allowing neutrons with energies down to 300~keV to be measured. 
As the most interesting region of the angular distribution is at backward
angles in the laboratory frame where the neutrons have the lowest energies,
this threshold can severely limit the PSD technique.  Liquid neutron
detectors can still be used in as TOF detectors below neutron energies of
about 1~MeV.\\

Ideally, these reactions would be performed in a configuration where the
detectors are shielded from both the neutrons and the $\gamma$ rays
emanating from the production target, the beam dump would not be directly
viewed by the detectors, and the room would have low levels of background
radiation.  In most cases it is necessary to tag on the reaction of
interest, usually by requiring an increase in charge of the recoil by one
unit compared to the beam species.  As the recoil cone is small and forward
focussed this requires using a detector system that has $Z$-discrimination and
can take the full beam rate.  Options include phoswich (phosphor sandwich
detectors), ionization chambers, magnetic recoil separators, and time-of-flight setups.  The choice of recoil detector depends on the specifics of
the experiment, namely the beam rate and the $\Delta$$Z$/$Z$ (i.e.  1/$Z$), the
latter of which is straightforward for light nuclei, but becomes
increasingly stringent for heavier elements.\\

Since these first attempts at $^7$Be(d,n)$^8$B and $^{17}$F(d,n)$^{18}$Ne
measurements, there have been significant upgrades to TWINSOL, with more
shielding between the source and the experimental area \cite{OMa16}.  At the
same time, the production efficiency has been increased by reducing the
thickness of the gas cell windows, and the ion optics have been improved to
provide a smaller beam spot.  Additionally, recoil identification detectors
will aid in channel selection to reduce background in future VANDLE
measurements with RIBs.  \\

\section{Inverse Kinematics (d,n) reactions with stable beams}

A stable beam experiment was run at the Nuclear Structure Laboratory (NSL)
at the University of Notre Dame in preparation for future TWINSOL
experiments and to investigate the response of VANDLE to inverse kinematics
(d,n) reactions where signal-to-noise levels are more optimal than with RIBs
and the level of statistics is not limiting.  A total of 21 VANDLE detectors
were mounted, at a distance of 0.5~m from a deuterated polyethylene (CD$_2$)
target, at laboratory angles from 65$^{\circ}$ to 170$^{\circ}$.  The stop
signal for the time of flight (ToF) was provided by the delayed RF signal
from the sweeper/buncher after the FN tandem.  The bunches came at 100~ns
intervals.  The beams used were $^{12}$C, at 8 energy steps between 18.5 and
41.7~MeV, and $^{16}$O at 64~MeV.  A 413~$\mu$g/cm$^2$ CD$_2$ target was
used for both beams, a second, thicker 711 $\mu$g/cm$^2$ CD$_2$ target was
used for the $^{12}$C beam.  In total, approximately 2 hours of beam were
used for the 8 energy steps with the $^{12}$C beam, and an hour for the
$^{16}$O.  The beam intensities were 4 - 5~$e$nA in each case, which
represents about 1-5 $\times$ 10$^9$~pps.\\

\begin{figure}[ht]
\centerline{%
\hspace{-2cm} \includegraphics[trim = 0 40 200 170, clip,width=13cm]{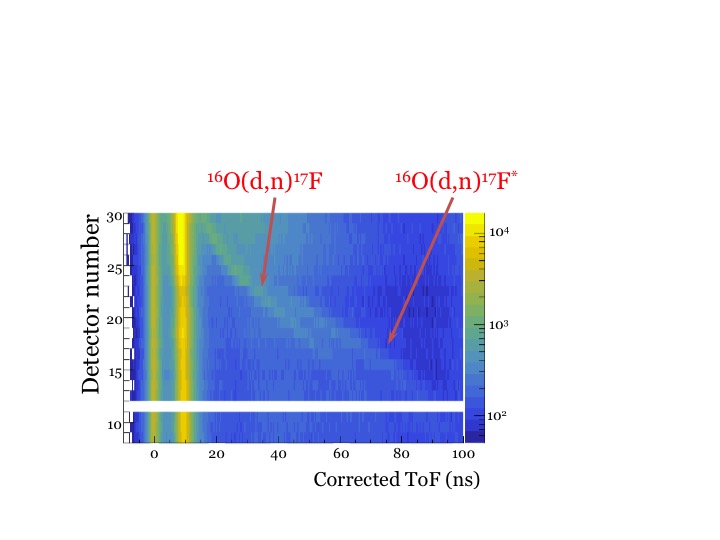}}
\caption{Detector number versus corrected time of flight for neutrons
following the $^{16}$O(d,n)$^{17}$F reaction, in inverse kinematics,  at
E$_{^{16}\rm{O}}$ = 64 MeV.  The polar angle in the laboratory frame
increases with decreasing detector number.  The detectors covered a range
from 65$^{\circ}$ to 170$^{\circ}$ in the laboratory frame.  The two
vertical lines at 0~ns and 10~ns are from the flashes of $\gamma$ rays
emitted as the beam struck the target and the beam stop, respectively.  }
\label{Fig:16odn}
\end{figure} 
\begin{figure}[htb]
\centerline{%
\hspace{0cm}\includegraphics[trim = 0 0 0 0, clip, width=12cm]{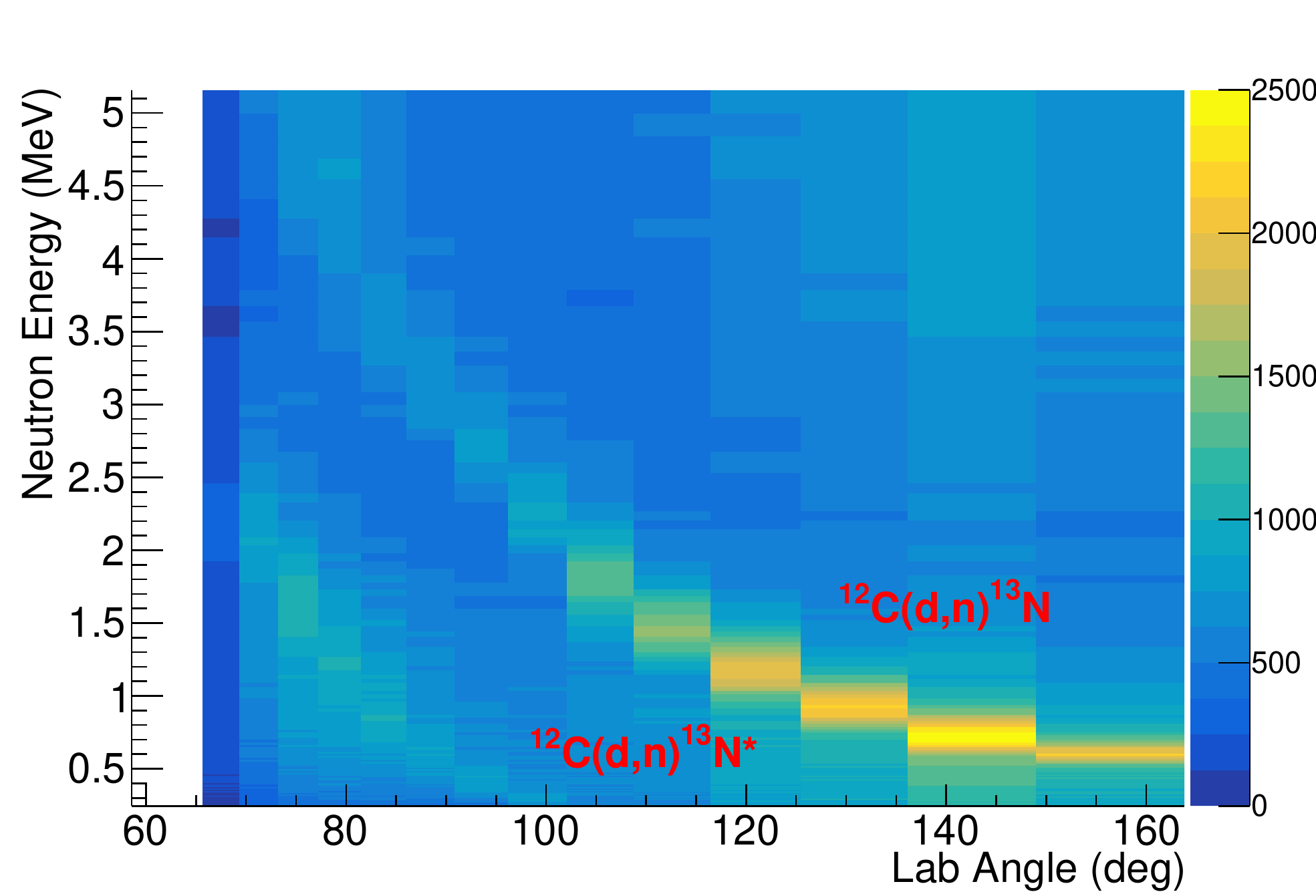}}
\caption{Angle versus energy plot for neutrons following the
$^{12}$C(d,n)$^{13}$N reaction, in inverse kinematics, at E$_{^{12}\rm{C}}$
= 41.7~MeV.  The loci show the population of the ground state and an excited
state at 2.365~MeV.  Smoothing related to the width of each detector was
applied in the calculation of the polar angle.}
\label{Fig:12cdn}
\end{figure}

Fig. \ref{Fig:16odn} shows the type of spectra seen online from just a few
minutes of beam, $^{16}$O in this case.  Each detector covered approximately
5$^{\circ}$ in the laboratory frame, with the most backward angles, from
170$^{\circ}$, at the bottom of the figure.  The two vertical lines, where
all detectors fired at the same ToF regardless of angle, are the $\gamma$
flashes from the target and the beam stop.  The target $\gamma$ flash was
used as a timing reference for measuring the kinetic energy of the neutrons. 
The two loci relating to the proton-transfer to the ground and first excited
states in $^{17}$F traverse the plot from the top left, close to the beam
stop $\gamma$ flash toward the bottom right, where the intensity fades to
around that of the background.\\

 Fig.  \ref{Fig:12cdn} shows the same information, but for the
$^{12}$C(d,n)$^{13}$N reaction, after the conversion of the neutron ToF
into energy and detector number into the laboratory angle.  The ground state
population is seen as a locus sweeping across from the top-center of the
plot toward the lower right.  As in inverse kinematics (d,p) measurements,
the emergent particle has higher energies at forward angles and low energies
at backward angles.  \\
 
\begin{figure}[htb]
\centerline{%
\includegraphics[width=\textwidth]{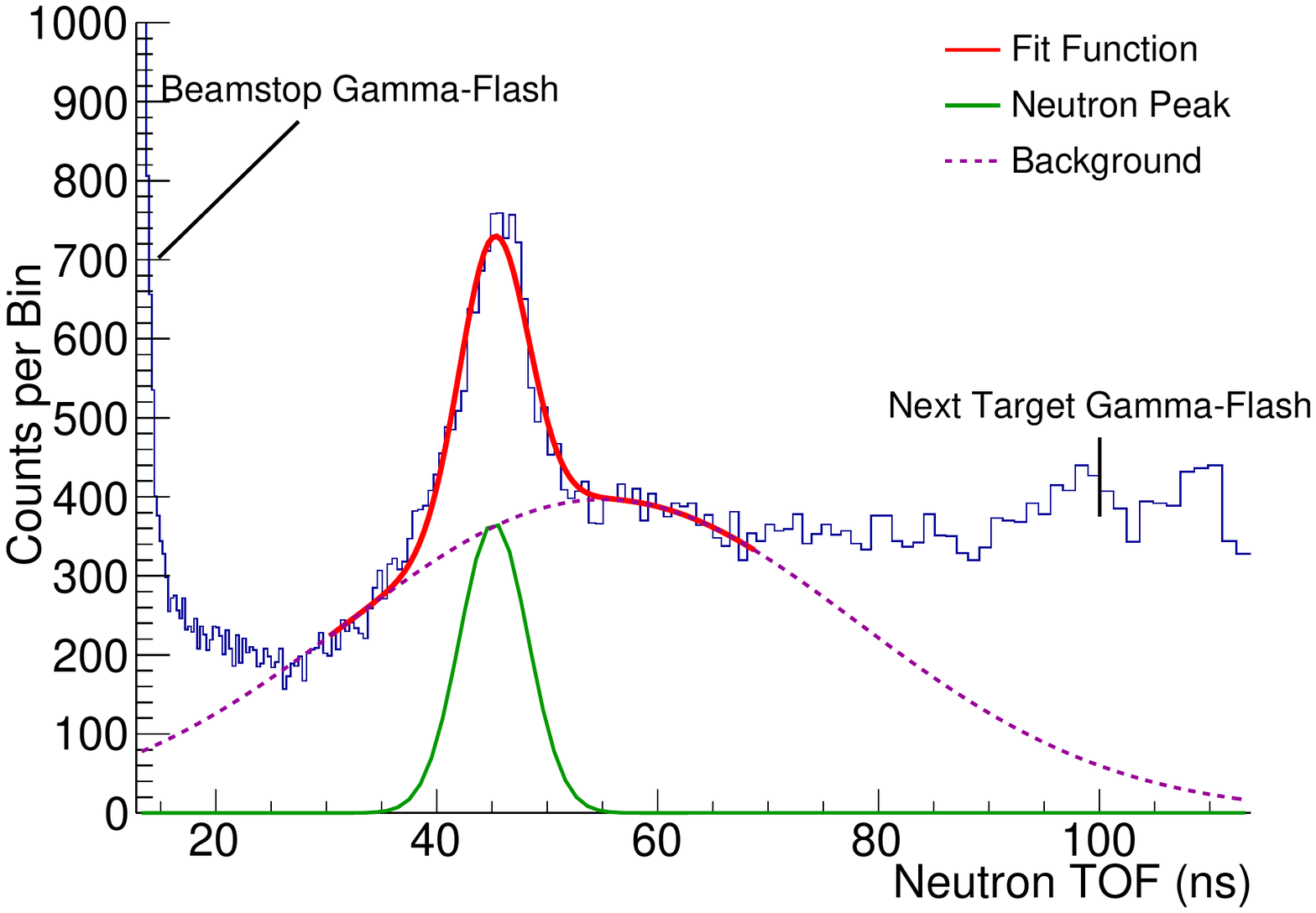}}
\caption{Neutron time-of-flight spectrum for the $^{12}$C(d,n)$^{13}$N
reaction at $\theta_{CM} = 12^{\circ}$ ($\theta_{lab} = 144^{\circ}$), in
inverse kinematics, with a beam energy of E$_{^{12}\rm{C}}$ = 41.7~MeV.  The
gaussian fit to the peak for population of the ground state, the background,
and the sum fit function are shown in green, magenta, and red respectively. 
The timing of the $\gamma$ flash from the next beam pulse is shown.}
\label{Fig:c12dnfit}
\end{figure}

 The data for each beam at each beam energy were subdivided by polar angle
of the neutron in the center-of-mass frame.  The intensity of each state was
found by fitting a gaussian curve on a linear background, as shown in Fig.
\ref{Fig:c12dnfit}.  These fits were made on the neutron ToF spectra as the
resolution is linear, as apposed to neutron energy spectra where the
resolution goes as ToF$^2$.  These intensities can then be plotted versus
center-of-mass angle to produce angular distributions.  A normalization was
performed using the known target thickness and the beam intensity, monitored
by a current integrator connected to an electrically insulated brass beam
stop.  The differential cross sections are currently being finalized and are
beyond the scope of this contribution.  Those that have been analyzed are of
sufficient quality to make meaningful comparisons with previous data and
state-of-the-art theory.

\section{Outlook}

The (d,p) reaction has been used extensively in inverse kinematics since the
mid-1990's, thus allowing transfer reaction techniques to be performed on
beams of exotic nuclei.  This has proven to be a powerful method for
extracting spectroscopic information relating to single-neutron states away
from stability.  In order to study single-proton states in an analogous way
it is necessary to develop techniques relating to the (d,n) single-proton
transfer reaction.  Our collaboration performed the (d,n) reaction on beams
of stable $^{12}$C and $^{16}$O with the neutron ToF array VANDLE.  High
quality data have been extracted from these measurements and normalized
differential cross-sections are currently being finalized.  \\

Channel selection is important to VANDLE measurements with RIBs, where
backgrounds from both neutrons and $\gamma$ rays are often limiting. 
Currently we are developing a compact ionization chamber with an internal
scintillator to provide a tag on the $Z$ of the recoil.  The scintillator will
be instrumented with silicon photomultipliers.  The results presented here
from stable beam measurements are encouraging and provide motivation to
develop recoil particle identification detectors such that the (d,n)
reaction, measured with ToF detector arrays, can be used as a spectroscopic
tool with RIBs.

\section*{Acknowledgments}
This research was supported in part by the U.S. Department of Energy, Office of Science, Office of Nuclear Physics under Contract No. DE-FG02-96ER40963 (UT), DE-AC05-00OR22725 (ORNL), by the National Science Foundation under Contract No. PHY1713857 (ND), PHY0354870 and PHY1404218 (Rutgers).  This research was sponsored in part by the National Nuclear Security Administration under the Stewardship Science Academic Alliance program through DOE Cooperative Agreement No. DE-NA0002132 (Rutgers and UTK).  STM would like to acknowledge funding from the Louisiana State University Department of Physics and Astronomy.

\end{document}